\documentclass[useAMS,usenatbib, referee]{mn2e}

\newcommand{\be}{\begin{equation}}
\newcommand{\ee}{\end{equation}}

\title[Spindown of massive rotating stars]{Spindown of massive rotating stars}
\author[H. B. Lau, A. T. Potter \& C. A. Tout]{Herbert H. B. Lau$^{1,2,3}$\thanks{E-mail:
hblau@astro.uni-bonn.de}, Adrian T. Potter$^2$ and Christopher A. Tout$^{2,1}$\\
$^1$Centre for Stellar and Planetary Astrophysics, School of Mathematics,
Building 28, Monash University,\\
Clayton VIC 3800, Australia\\
$^2$Institute of Astronomy, The Observatories, Madingley Road, Cambridge CB3 0HA
$^3$Argelander Institute for Astronomy, University of Bonn, Auf dem Huegel 71, D-53121 Bonn, Germany\\
}

\begin{document}
\date{Accepted 2011 March 21}

\pagerange{\pageref{firstpage}--\pageref{lastpage}} \pubyear{0000}

\maketitle

\label{firstpage}

\begin{abstract}
Models of rapidly rotating massive stars at low metallicities show
significantly different evolution and higher metal yields compared to
non-rotating stars. We estimate the spin-down time-scale of rapidly
rotating non-convective stars supporting an $\alpha-\Omega$
dynamo. The magnetic dynamo gives rise to mass loss in a magnetically
controlled stellar wind and hence stellar spin down owing to loss of
angular momentum. The dynamo is maintained by strong horizontal
rotation-driven turbulence which dominates over the Parker
instability. We calculate the spin-down time-scale and find that it
could be relatively short, a small fraction of the main-sequence
lifetime. The spin-down time-scale decreases dramatically for higher
surface rotations suggesting that rapid rotators may only exhibit such
high surface velocities for a short time, only a small fraction of
their main-sequence lifetime.
\end{abstract}

\section{Introduction}
Models of rotating stars, particularly those of low-metallicity,
have generated a great deal of interest because it is believed that
their nucleosynthetic output could be significantly affected and so may
better fit the observed composition of halo stars and globular
clusters.  Otherwise unaccountable nitrogen in very metal-poor
stars could be explained by rapidly rotating massive stars because
their nitrogen yields are found to greatly increase with spin
\citep[e.g.][]{meynet2006}.  There are suggestions that stars of very low
metallicity could have very high equatorial spin velocities, of the order
of $600-800\,\rm km \, s^{-1}$ \citep{chiappini2006}.  Moreover very
rapidly rotating stars could induce chemically homogeneous evolution
and this may solve a problem of insufficient mass loss at very low
metallicity in the collapsar models of $\gamma$-ray bursts \citep{yoon2005}.

\citet{tout1991} showed how a magnetic dynamo might be driven by
rapid rotation and radial turbulence. The mass loss in a magnetically
controlled stellar wind results in quite rapid angular momentum loss
and hence stellar spin down. In this paper we present an estimate of
the spin-down rate for non-convective stars. In all cases the Alfv\'en
radius is larger than the stellar radius so angular momentum may
be efficiently removed from the system and this leads to short
spin-down time-scales.

Recently, \citet{dervisoglu2010} considered the spin angular momentum
evolution of the accreting components of Algol-type binary stars and
demonstrated how accretion from a disc ought to spin the accreting
star up to breakup velocity very quickly, after only a small amount of
mass transfer.  However, the accreting stars in Algols, which are hot
main-sequence stars with radiative envelopes are not spinning at
anywhere near their breakup rate.  They showed that tides alone are
insufficient to explain the spin down and argued that the most likely
mechanism for the angular momentum loss is a magnetically controlled
wind generated by a dynamo which is itself driven by the rapid
rotation.  This is further evidence that stars cannot continue to spin
close to their breakup velocity and that a magnetic dynamo is a likely
the spin-down mechanism.

\section{Spin-down time-scale and magnetic wind}

We use as a basis the model of \citet[][ hereinafter TP]{tout1991}.
In their model mass loss and braking are powered by an $\alpha-\Omega$
dynamo that extracts energy from differential rotation. The
differential rotation at the surface and core boundary is maintained
by stellar winds and core contraction respectively. Differential
rotation throughout the radiative envelope results from redistribution
of angular momentum because of the meridional circulation. This can be
seen most clearly in models where angular momentum transport is
treated as a purely diffusive process. In these models there is very
little differential rotation \citep{potter2011}. In the case where
there is a strong magnetic field, the meridional circulation may be
inhibited \citep{maeder2003b} but this conclusion has been questioned
by \citet{maeder2005} who found that, if the meridional circulation is
neglected, there is insufficient differential rotation to maintain the
magnetic dynamo. In particular, if differential rotation is too
strongly suppressed by magnetic fields then it is difficult to explain
the existence of slow rotating stars with strong nitrogen enrichment
\citep{hunter2009, meynet2010}.  Mixing may be driven either by
Parker's instability or by the horizontal rotation-driven turbulence
that is itself responsible for the rotationally driven mixing.  In
TP's model convection served this purpose.  Otherwise our
$\alpha-\Omega$ dynamo operates in a similar way.  Here we find that
the extremely strong horizontal turbulence dominates over the Parker
instability, even for low rotation rates.

Our calculations are similar to those of TP.  For a star of mass $M$,
radius $R_{\ast}$ and angular velocity $\Omega$, the spin-down
time-scale $\tau_{\rm sd}$ is the total angular momentum divided by
the rate of angular momentum loss
\begin{equation}
\tau_{\rm sd} \approx \frac {k^{2}MR_{\ast}^{2} \Omega}{\dot{J}_{\rm w}},
\end{equation}
where $k$ is the dimensionless stellar radius of gyration ($k^{2}
\approx 0.1$) and $\dot{J}_{\rm w}$ is the angular momentum loss rate
in the wind.
\begin{equation}
\dot{J}_{\rm w}\approx \dot{M}_{\rm w} R_{\rm A}^{2} \Omega , 
\end{equation}
where $R_{\rm A}$ is the Alfv\'en radius \citep{mestel1968} where the magnetic energy and
kinetic energy in the plasma balance.
At $R_{\rm A}$
\begin{equation}
\label{vwind}
v_{\rm w}^{2} \approx B^{2}/\mu_0\rho_{\rm w}
\end{equation}
where $\rho_{\rm w}$ is the density of wind material and B is the
magnetic flux density.  We expect the velocity of the wind $ v_{\rm
w}$ is of the order of the escape velocity so
\begin{equation}
\label{escapevel}
v_{\rm w}^{2} \approx (2GM/R_{\ast}).
\end{equation}
We are imagining a wind launched by magnetic heating of a hot
corona and accelerated by coronal pressure here.  This differs from
the radiatively driven winds normally associated with massive stars
that would be somewhat faster.  We have also assumed that the
wind is not significantly slowed by the star's gravity.
To quantify the effect of changing $v_{\rm w}$ we write
\begin{equation}
v_{\rm w}^{2} \approx \beta(2GM/R_{\ast}),
\end{equation}
where $\beta > 1$ if the wind is accelerated and $\beta < 1$ if
it is slowed.  Though we expect $R_{\rm A} > R_*$, if it does fall
below $R_*$ then the angular momentum loss rate is instead
\begin{equation}
\dot{J}_{\rm w}\approx \dot{M}_{\rm w} R_{\ast}^{2} \Omega 
\end{equation}
because the material is then lost with the specific angular momentum
of the stellar surface.

Outside the star we assume the field strength $B$ depends on radius as
\begin{equation}
\label{bfield}
B=B_{\ast}{(R_{\ast}/R)}^{n},
\end{equation}
where $B_{\ast}$ is the field at the surface of the star.
The outflow
at the Alfv\'en surface is roughly spherical so that
\begin{equation}
\label{density}
\rho_{\rm w} \approx \dot M_{\rm w}/(4\pi R_{\rm A}^{2} v_{\rm w})
\end{equation}
and the relation between $R_{\rm A}$ and $R_{\ast}$ follows the
equations~(\ref{vwind}), (\ref{bfield}) and~(\ref{density}).
\begin{equation}
\frac{R_{\rm A}}{R_{\ast}} =  {\left(\frac {4\pi B_{\ast}^{2} R_{\ast}^{2}}
{\mu_0\dot M_{\rm w} v_{\rm w}}\right)}^{1/(2n-2)}.
\end{equation}
An undisturbed dipole field has $n=3$.  If the wind were able to
open the field lines, so that $n=2$, then $R_A \ge R_*$ would be
larger and the spindown correspondingly faster.

To estimate the wind mass-loss rate $ \dot M_{\rm w}$ we consider the
rate at which magnetic energy is dissipated and assume that all of
this energy $L_{\rm w}$ goes into launching the wind
\begin{equation}
L_{\rm w} \approx \alpha \frac{GM\dot{M}_{\rm w}} {R_{\ast}}.
\end{equation}
We include the factor $\alpha$ because it is likely that not all
the magnetic energy released is available to launch the wind.  It
might simply be radiated away.  In that case $\alpha > 1$.  On the
other hand it might even be possible that less energy is needed to
drive the wind because it may be accelerated by radiation driving
but only at high metallicity.  In such a case $\alpha < 1$.

TP estimated the differential rotation by assuming that convection
efficiently redistributes angular momentum.  In non-accreting,
non-convective, rotating stars the strength of the differential
rotation results from the relative strength of meridional circulation
and turbulent diffusion. The latter tends to smooth out differential
rotation whereas the former in general enhances it. Turbulent
diffusion may arise owing to purely hydrodynamic or magnetic
instabilities. The exact formulation with and without magnetic fields
has been the subject of much discussion and many different models are
commonly used \citep[for examples see][]{heger2000, meynet2000, maeder2004,
meynet2010}.  In all cases, the energy for driving the dynamo comes
essentially from the excess rotational kinetic energy in the shear,
$E_{\rm shear} = \frac {1}{2} k^{2} MR_{\ast}^{2} \Delta\Omega^2$,
where $\Delta\Omega$ approximates the range of angular velocity across
the stellar radius.  The rate $L_{+}$ at which energy is fed into the
shear is
\begin{equation}
L_{+} \approx \frac{\frac {1}{2} k^{2} MR_{\ast}^{2}
\Delta\Omega^{2}}{\tau_{\nu}},
\end{equation}
where $\tau_{\nu}\approx\frac{R_{\ast}^2}{\nu}$ is the viscous time
scale and $\nu$, which behaves as a viscosity, is the radial diffusion
coefficient for angular momentum.

We are considering radial variations of the rotation rate
under a shellular framework \citep{zahn1992} so it is appropriate to take the
radial diffusion coefficient for Kelvin-Helmholtz
instabilities driven by the shear, referred to as $D_{\rm shear}$ by
\citet{meynet2000}.  Thus $\Delta\Omega\approx\frac{\partial\Omega}{\partial\ln
r}\vline_{\,r_0}$ is a measure of differential rotation in the
star.  Because $\frac{\partial\Omega}{\partial\ln r}$ varies throughout the
star we choose $r_0$ to give a typical value within the radiative
envelope.  We then evaluate the diffusion coefficients at the same
radius.  For massive stars, the amount of differential rotation is
governed by the rate of evolution and the strength of the meridional
circulation driven by rotation. Assuming that hydrostatic evolution is slow compared
to the advection and diffusion time-scales for angular momentum we may use 
\citet{zahn1992}'s model in steady state to get
\begin{equation}
\Delta\Omega\approx \frac{r U \Omega}{5\nu},
\end{equation}
where $U P_2(\cos\theta)$ is the radial component of the meridional
circulation and $P_2(x)$ is the second Legendre polynomial in $x$.
Using the formulation of \citet{talon1997} for the
shear viscosity which dominates the behaviour of the radial
differential rotation \citep{meynet2000} we can approximate this as
\begin{equation}
\Delta\Omega\approx \left(\frac{2 r U \Omega N_{\rm T}^2}{K}\right)^{\frac{1}{3}},
\end{equation}
where $N_{\rm T}^2$ is the Brunt--V\"ais\"al\"a frequency and $K$ is
the thermal diffusivity. It has been shown that the stationary
approximation to the dynamical equation of \citet{zahn1992} is quite
poor \citep{meynet2000}.  However this is in the case where the
meridional circulation is inferred from the steady state differential
rotation. Here we are using the meridional circulation and turbulent
diffusivity calculated from non-stationary models to parameterize the
differential rotation so we expect the approximation to be much better
than a purely stationary model.
We assume that the magnetic field strength does
not become strong enough to cause dynamical feedback on either the
rotation or the meridional circulation.

Assuming that the energy dissipated from the shear is fed into driving
a stellar wind, $L_+\approx L_{\rm w}$, we may derive the spin-down
time-scales
\begin{equation}
\label{SD2}
\tau_{\rm
  sd}\approx\frac{k^2MR_*^2\Omega}{\dot{M}_{\rm
    w}[\max (R_*,R_{\rm A})]^2\Omega}\approx
\cases{\alpha^{1/2}\beta^{1/4}
k\left(\frac{2GM}{R_*^3}\right)^{\frac{3}{4}}\Delta\Omega^{-1}
\tau_{\nu}^{\frac{1}{2}}\left(\frac{\mu_0M}{4\pi B_*^2R_*}\right)^{\frac{1}{2}}
& $R_{\rm A}>R_{\ast}$\cr
\alpha\left(\frac{2GM}{R_{\ast}^3}\right)\tau_\nu\Delta\Omega^{-2} & $R_{\rm A}<R_{\ast}$.}
\end{equation}
The driver of the stellar dynamo is the shear, $\Delta\Omega$, which
converts the poloidal field $B_{\rm p}$ to toroidal $B_{\phi}$.
Cyclonic turbulence can then regenerate poloidal flux from toroidal
\citep{parker1955,cowling1981}, so we have
\begin{equation}
\label{bphi}
\frac {d B_{\phi}} {dt} \approx \Delta \Omega B_{\rm p} - \frac
{B_{\phi}} {\tau_{\phi}}
\end{equation}
and
\begin{equation}
\label{bp}
\frac {d B_{\rm p}} {dt} \approx (\frac{\Gamma}{R_{\ast}}) B_{\phi} -
\frac {B_{\rm p}} {\tau_{\rm p}},
\end{equation}
where $\Gamma$ is the dynamo regeneration term and $\tau_{\rm p}$ and
$\tau_{\phi}$ are the time-scales on which poloidal and toroidal flux
are lost or destroyed.  Following TP we take $\tau_{\rm p} \approx
\tau_{\phi} \approx 10 \tau_{\rm A}$, where $ \tau_{\rm A} \approx
R_{\ast}/v_{\rm A}$ is the Alfv\'en-wave crossing time in the star for
an Alfv\'en speed $v_{\rm A}$.  As did TP, we write
\begin{equation}
\label{gamma}
\Gamma \approx \gamma v_{\rm t},
\end{equation}
where $\gamma$ is an unknown parameter which describes the efficiency
of the regeneration term and $v_{\rm{t}}$ is a characteristic
turbulent-eddy velocity.  The factor $\gamma$ is very uncertain.
Based on the evolution of cataclysmic variables \citep{warner1995} TP
set it to $10^{-2}$ for convective stars but found a much lower value
of $3\times 10^{-5}$ to be sufficient to drive the activity in Ae and
Be stars \citep{tout1994}.  We use a conservative estimate of $\gamma
= 10^{-4}$ here and later show that our estimated spin-down rate is
proportional to $\gamma$.

In a steady state equations~(\ref{bphi}) and~(\ref{bp}) lead to
\begin{equation}
\label{bratio}
B_{\rm p} \approx \frac{\tau_{\rm p} \Gamma}{R_{\ast}}B_{\phi}
\approx 0.1\frac{v_{\rm A}}{\Delta\Omega R_{\ast}}B_{\phi}.
\end{equation}
For each of the models calculated $B_{\phi} \approx 10^3 B_{\rm p}$ so
that
\begin{equation}
B_{\phi} \approx v_{\rm A} \sqrt{\mu_0\rho_{\ast}},
\end{equation}
where
\begin{equation}
\label{meandensity}
\rho_{\ast} = M/(\mu_0R_{\ast}^{3}/3)
\end{equation}
is the mean density of the star. Equation~(\ref{bratio}) then gives
\begin{equation}
B_{\rm p}\approx\frac{0.1}{\Delta\Omega}\frac{v_{\rm A}^2\sqrt{4\pi\rho_*}}{R_*}
\end{equation}
and
\begin{equation}
\label{alfven}
v_{\rm A}^2\approx 100R_*\Gamma\Delta\Omega.
\end{equation}
We approximate the eddy velocity of turbulence owing to shear by
$v_{\rm{t}}\approx \max (v_\nu,0.1v_{\rm A})$ where $v_\nu$ is the
characteristic velocity of rotation-driven eddies.  So either the
rotation driven turbulence dominates the dynamo regeneration or, if it
is weak enough, the Parker instability sets in.  TP set
$v_\nu^2\approx 3D_{\rm h}\Omega$, where $D_{\rm h}$ is the horizontal
diffusion coefficient.  When rotationally driven turbulence dominates
$D_{\rm h}\gg\nu$ and is principally responsible for driving the
dynamo.  We find this to be the case in all models we consider below
and so do not consider the alternative in any detail.

It is the poloidal component of the magnetic field that predominantly
drags around the wind so equations~(\ref{gamma})
and~(\ref{meandensity})~--~(\ref{alfven}) lead to
\begin{equation}
B_{\ast} \approx B_{\rm p} \approx 10\sqrt{3\mu_0/4\pi}\gamma
R_*^{-\frac{3}{2}}M^{\frac{1}{2}} v_{\rm{t}}.
\end{equation}
We define time-scales
\begin{equation}
\tau_\nu = \frac{R_*^2}{\nu},
\end{equation}
\begin{equation}
\tau_{\rm shear} = \Delta\Omega^{-1}
\end{equation}
and
\begin{equation}
\tau_{\rm dyn} = \left(\frac{R_*^3}{2GM}\right)^\frac{1}{2}
\end{equation}
so that
\begin{equation}
\dot M_{\rm w} = \frac{\tau_{\rm dyn}^2}{\tau_{\rm shear}^2}\frac{Mk^2}{\alpha\tau_\nu}
\end{equation}
and
\begin{equation}
v_{\rm w} = \beta^{1/2}\frac{R_*}{\tau_{\rm dyn}}
\end{equation}
then with
\begin{equation}
\tau_{\rm turb} = \frac{R_*}{v_{\rm t}}.
\end{equation}
We then have the ratio of the Alfv\'en to the
stellar radius,
\begin{equation}
\frac{R_{\rm A}}{R_{\ast}}\approx
\sqrt{10\sqrt{\frac{3\mu_0}{4\pi}}}\frac{\gamma^{1/2}\alpha^{1/4}}{k^{1/2}\beta^{1/8}}\left(\frac{\tau_{\nu}\tau_{\rm{shear}}^2}{\tau_{\rm{turb}}^2\tau_{\rm{dyn}}}\right)^{\frac{1}{4}},
\end{equation}
and a spin-down time-scale,
\begin{equation}
\label{SD1}
\tau_{\rm sd}\approx \cases{\frac{\alpha^{1/2}\beta^{1/4}k}{10\gamma\sqrt{3\mu_0/4\pi}}{\displaystyle\frac{\tau_{\rm{turb}}\tau_{\rm
shear}\tau_\nu^{\frac{1}{2}}}{\tau_{\rm dyn}^{3/2}}} & $R_{\rm A}>R_{\ast}$\cr
{\displaystyle\frac{\tau_{\rm shear}^2\tau_\nu}{\tau_{\rm dyn}^2}} & $R_{\rm A}<R_{\ast}$.}
\end{equation}

\section{Application to rotating massive stars}

\begin{table*}
\caption{
\label{spindown}
Properties of a $20\,M_\odot$ star spinning with equatorial velocity
$v$ (first column). The mass loss rate $\dot{M}/{\rm M_{\odot}\,yr^{-1}}$ is the predicted mass loss rate driven by the magnetic wind. The second column lists the ratio of the Alfv\'en
radius to the stellar radius.  This ratio is proportional to
$\alpha^{1/4}\beta^{-1/8}\gamma^{1/2}$.  The star's surface magnetic field strength is $B_*$
and typical radial and horizontal diffusion coefficients are $\nu$ and
$D_{\rm h}$.  Then $\Delta\Omega$ is an estimate of the differential
rotation through the star, $\tau_{\rm ms}$ is its main-sequence
lifetime, were it to continue to conserve its angular momentum, and
$\tau_{\rm sd}$ is its spin-down time-scale which is proportional to
$\alpha^{1/2}\beta^{1/4}\gamma^{-1}$.}
\begin{tabular}{ccccccccc}
\hline
$v/\rm km\,s^{-1}$ & $R_{\rm A}/R_{\ast}$ & $B_{\ast}/$G &
$\nu/\rm{cm^2\,s^{-1}}$ & $D_{\rm h}/\rm{cm^2\,s^{-1}}$ & $\dot{M}/{\rm M_{\odot}\,yr^{-1}}$ &
$\Delta\Omega/\rm{s^{-1}}$ & $\tau_{\rm ms}/{\rm yr}$ & $\tau_{\rm sd}/{\rm yr}$ \\
\hline
  50  & 2.1 & 14 & $1.0\times 10^8$ & $4.5\times 10^{12}$ & $2.7\times 10^{-10}$ & $2.3\times 10^{-5}$ & $7.8\times 10^6$ & $1.7\times 10^9$ \\
  100 & 1.7 & 37 & $3.3\times 10^8$ & $1.4\times 10^{13}$ & $3.9\times 10^{-9}$ & $4.8\times 10^{-5}$ & $7.9\times 10^6$ & $1.8\times 10^8$ \\
  150 & 1.5 & 62 & $7.0\times 10^8$ & $2.5\times 10^{13}$ & $1.9\times 10^{-8}$ & $7.4\times 10^{-5}$ & $8.1\times 10^6$ & $4.6\times 10^7$ \\
  200 & 1.4 & 97 & $1.0\times 10^9$ & $4.2\times 10^{13}$ & $5.3\times 10^{-8}$ & $9.9\times 10^{-5}$ & $8.3\times 10^6$ & $1.8\times 10^7$ \\
  300 & 1.3 & 160 & $2.0\times 10^9$ & $6.6\times 10^{13}$ & $2.2\times 10^{-7}$ & $1.5\times 10^{-4}$ & $8.9\times 10^6$ & $5.6\times 10^6$ \\
  400 & 1.2 & 220 & $2.6\times 10^9$ & $9.4\times 10^{13}$ & $5.1\times 10^{-7}$ & $2.0\times 10^{-4}$ & $9.4\times 10^6$ & $2.7\times 10^6$ \\
  500 & 1.2 & 281 & $3.0\times 10^9$ & $1.2\times 10^{14}$ & $9.0\times 10^{-7}$ & $2.5\times 10^{-4}$ & $9.7\times 10^6$ & $1.6\times 10^6$ \\
\hline
\end{tabular}
\end{table*}

To investigate the spin down of a $20\,M_\odot$ star we evaluated the
spin-down time-scale (equation~\ref{SD1}), with $\alpha = 1$, $\beta =
1$ and $\gamma = 10^{-4}$, for a series of surface rotation rates.  We
used the Cambridge STARS code \citep{eggleton1971}, which has
previously been updated by many authors \citep[most importantly
by][]{Pols1995,Stancliffe2009}.  The version we have used has been
modified to include the effects of rotation by \citet{potter2011}.
With this we can estimate $U$, $D_{\rm h}$ and $\nu$ for the star on
the main sequence.  In this case the Alfv\'en radius is always larger
than the radius of the star.  We chose to estimate the turbulence at
$r_0$ where $\nabla-\nabla_{\rm ad}$ has a minimum.  In all cases
$r_0\approx 4\,R_{\odot}$ from the centre of the star which has a
total radius $R_*\approx 6\,R_\odot$.  In Table~\ref{spindown} we list
various properties of the models, including the spin-down time-scale.
We note that $R_{\rm A}/R_*
\propto\alpha^{1/4}\beta^{-1/8}\gamma^{1/2}$ and $\tau_{\rm sd}\propto
\alpha^{1/2}\beta^{1/4}\gamma^{-1}$.  The mass-loss rate driven by
this mechanism is at a rate of $\dot M_{\rm w} \approx
\alpha^{-1}\,10^{-6}\,{\rm M}_{\odot}\,{\rm yr}^{-1}$ for the most
rapid rotators and escapes at $v_{\rm w} \approx
\beta^{1/2}\,10^3\,\rm km\,s^{-1}$.  This is a slower, but not much
slower, wind than that typically driven by radiation, at
about 3,000\,km\,$\rm s^{-1}$.  So we may expect $\beta \le 3$ and
this to have little effect. Also there is no a priori reason why
such a magnetically driven wind, or its consequences, should depend on
metallicity.  The spin-down time-scale becomes shorter than the
main-sequence lifetime for stars initially rotating faster than around
$200\,\rm km\,s^{-1}$.  It is much shorter for rapid rotators and so a
star cannot continue to rotate at a high velocity throughout its
entire main-sequence lifetime.  It appears unlikely that homogeneous
evolution, induced by very rapid rotation, can last long enough to
change the fate of the stars even when they are of low metallicity.
We note that $\zeta$ Puppis, a $53.9\rm M_{\odot}$ star
\citep{Repolust2004} with an observed rotational velocity of $v\sin i
\approx 200\rm{km s^{-1}}$ \citep*{Kudritzki1983}, has an observed
mass loss rate of about $\dot M_{\rm w} \approx 3.5\times 10^{-6}\,\rm
M_{\odot}\,yr^{-1}$ \citep{Cohen2010}. This mass loss rate is not
inconsistent with our estimate.

\subsection{Uncertainty in the estimation}

We have assumed perfect efficiency for the transfer of energy from
shear to the magnetically driven wind. This is unlikely to be the case
and so the spin-down time-scales are likely to be larger than those
predicted.  However, if the efficiency factor $\alpha$ does not depend on the
rotation rate, the prediction that rapid rotators spin down on a much
shorter time-scale is still valid. Indeed, unless the efficiency is
extremely low ($\alpha \ge 30$), the spin-down time-scale for rapid rotators is
much shorter than their main-sequence lifetime.

There is also uncertainty introduced by our choice of $r_0$.  We have
chosen it by a
standardised method within the radiative
envelope.  However, if we vary $r_0$ for a $20\,M_{\odot}$ star with surface rotation of
$200\,\rm km\,s^{-1}$, the spin-down time-scale varies between
$7.2\times 10^4$ and $2.1\times 10^7\,$yr.  It is unlikely to
be close to the lower limit because this is the spin-down time-scale
when $r_0$ is taken close to the convective core.  This is
only representative of a small part of the radiative envelope and energy
deposited there is unlikely to be very influential in driving the stellar
wind.

A major uncertainty is $\gamma$, the standard dynamo
regeneration term.  In our calculation, we have used a conservative
value of $10^{-4}$ but it could be larger by as much as two orders of
magnitude.  If that were the case then, the spin-down time-scales
($\tau_{\rm sd}\propto \gamma^{-1}$) would
be much shorter.  Given the uncertainties in our estimate, we cannot
draw a quantitative conclusion for a star with a moderate spin
rate.  However, our results do suggest that stars rotating near their
break-up velocity, or stars with very efficient rotationally driven
mixing, are likely to spin down very quickly by driving a magnetically
locked wind. This effect cannot be ignored.

As table \ref{spindown} shows, $R_A>R_*$ in all of the cases examined
here. This means that $\tau_{\rm sd}\propto\Delta\Omega$, assuming
that the turbulent viscosity is unaffected. If the differential
rotation is strongly suppressed by the presence of a magnetic field
then it may extend the spindown timescale of the star. However, the
degree to which this happens is extremely uncertain. In particular, if
solid body rotation is enforced it is difficult to reconcile models
and observations \citep{meynet2010}.

\section{conclusion}

We have estimated the spin-down time-scale for rapidly rotating
non-convective stars.  With a conservative value for the magnetic
field regeneration term $\gamma$, the spin-down time-scale is likely
to be long for slowly rotating stars for which the horizontal turbulence
driven by rotation is weaker.  However, for more rapid rotators, the
spin-down time-scale can be much shorter than the main-sequence
lifetime.  Given the uncertainties in the coefficient of mixing,
viscosity and the appropriate mixing length, more detailed models of
magnetic spin down driven by rotational mixing must be developed
before we can decide on whether rapid spin can lead to homogeneous
mixing and on the importance of rotationally driven mixing at lower
spin rates. These calculations do however mean that we should
  not automatically assume that even very low-metallicity stars can
  continue to spin rapidly throughout their main-sequence lives.

\section{Acknowledgements}

HBL and CAT thank Dayal Wickramasinghe and ANU for hospitality when
finishing this work.  CAT thanks Churchill college for his fellowship.
ATP thanks the STFC for his studentship. We thank the anonymous referee for useful comments to improve manuscript.

\label{lastpage}

\end{document}